\documentclass{elsart}
\usepackage[dvips]{epsfig}

\begin{document}
\begin{frontmatter}

\title{Subdiffusion and the cage effect studied near the
colloidal glass transition}
\author[emory]{Eric R.~Weeks\corauthref{cor1}},
\corauth[cor1]{Corresponding author.}
\ead{weeks@physics.emory.edu}
\author[harvard]{D.~A.~Weitz}
\address[emory]{Physics Department, Emory University, Atlanta, GA 30322, USA}
\address[harvard]{DEAS and Physics Department, Harvard 
University, Cambridge, MA 02138, USA}

\begin{abstract}
The dynamics of a glass-forming material slow greatly near the
glass transition, and molecular motion becomes
inhibited.
We use confocal microscopy to investigate the motion of colloidal
particles near the colloidal glass transition.  As the
concentration in a dense colloidal suspension is increased,
particles become confined in transient cages formed by their
neighbors.  This prevents them from diffusing freely throughout
the sample.  We quantify the properties of these cages by
measuring temporal anticorrelations of the particles'
displacements.  The local cage properties are related to the
subdiffusive rise of the mean square displacement:  over a broad
range of time scales, the mean square displacement grows slower
than linearly in time.
\date{\today}
\end{abstract}
\begin{keyword}
Glass transition \sep colloids \sep anomalous diffusion
\sep cage effect

\PACS 61.43.Fs \sep 64.70.Pf \sep 82.70.Dd \sep 05.40.Fb
\end{keyword}
\end{frontmatter}

\section{Introduction}

As glass-forming materials are cooled, the sharply increasing viscosity
of the liquid is accompanied by equally dramatic
changes in the motion of tracer particles within the material
\cite{reviews}.
In particular, the mean square displacement $\langle \Delta x^2
\rangle$ (MSD) of an ensemble
of tracer particles embedded in a glass-forming material
forms a plateau at intermediate lag times, reflecting the
crowding of the particles which prevents easy rearrangements
[see Fig.~\ref{time1}(a)].  At longer lag times, the MSD
shows an upturn, returning to diffusive
motion, albeit with a greatly reduced diffusion coefficient
($\langle \Delta x^2 \rangle \sim 2 D_{\infty} \Delta t$).
Dense colloidal suspensions are simple materials which undergo
a glass transition as the particle concentration increases, and
provide a way to directly study the anomalous kinetics of the
colloidal particles near the glass transition, to determine how
the local motion of individual particles gives rise to the unusual behavior
of the ensemble MSD \cite{weeks00,kegel00}.
The plateau in the MSD is subdiffusive: for
a range of time scales $\Delta t$, $\langle \Delta x^2 \rangle$
grows as $\langle \Delta x^2 \rangle \sim (\Delta t)^\gamma$ with
$\gamma < 1$; $\gamma=1$ is the more typical diffusive case.
Typically subdiffusion arises when a system possesses memory
\cite{weeks98}.
In this work, we test this by looking for temporal correlations
in the particle motions.  We find that these correlations do
exist and are due to the ``cage effect'' of glassy systems
(see Fig.~\ref{trajs}).  We characterize this cage effect,
and directly connect the local description of particle caging
to the subdiffusive plateau in the MSD and
the lag-time dependent anomalous diffusion exponent $\gamma
(\Delta t)$.


\section{Experimental procedure}

Our samples are colloidal poly-(methylmethacrylate)
(PMMA) particles, sterically stabilized by a thin layer of
poly-12-hydroxystearic acid \cite{weeks00,dinsmore,pmma}.
They are in an organic solvent mixture of cyclohexylbromide
and decalin, chosen to closely match the density and index of
refraction of the particles \cite{dinsmore}.  The particles have
a radius $a=1.18$~$\mu$m and a polydispersity of $\sim$5\%.
They are dyed with rhodamine dye, which results in a slight
charging of the particles.  Despite this slight charge,
their phase behavior is similar to colloidal hard spheres
\cite{pusey86}:  we find $\phi_{\rm freeze}=0.38$ and $\phi_{\rm
melt}=0.42$ (for hard spheres these values are $\phi_{f}=0.494$
and $\phi_m=0.545$).  As the concentration is further increased,
we see a glass transition at $\phi_{\rm g}\approx 0.58$, in
agreement with what is seen for hard spheres.  Samples with
$\phi>\phi_g$ do not form crystals in the bulk even after
they have been sitting for several months.
Moreover, the diffusion constant for such
samples goes to zero -- the samples become nonergodic
\cite{reviews}.


We view the colloidal particles with a fast scanning laser
confocal microscope, to obtain three-dimensional
images from deep within the sample
\cite{weeks00,kegel00,dinsmore,alfons95}.  In practice, we focus
at least 30~$\mu$m from the coverslip of the sample chamber,
to avoid wall effects.  By taking a series of three-dimensional
images at intervals of 10-20~s, we are able to follow the motion
of several thousand colloidal particles for several hours.
We identify particle centers with an accuracy of 0.03~$\mu$m
horizontally and 0.05~$\mu$m vertically; the poorer vertical
resolution is due to optical limitations of the microscope.
For further details, see Refs.~\cite{dinsmore,crocker}.

\section{Results}

\begin{figure}
\centerline{
\epsfxsize=6.0truecm
\epsffile{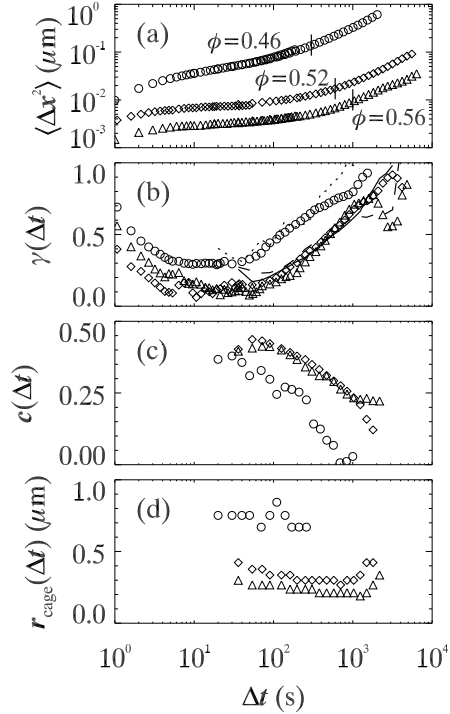}}
\smallskip
\caption{(a) Mean square displacement for three ``supercooled
fluids,'' with volume fractions $\phi$ as indicated.
The vertical lines indicate the cage
rearrangement time scale $\Delta t^*$.
(b) The symbols indicate the
measured anomalous diffusion exponent $\gamma(\Delta t)$, equivalent
to the logarithmic slope of $\langle \Delta x^2 \rangle$.
The lines show
the predicted value $\gamma_{\rm est}$ based on Eq.~\protect\ref{slope} (dotted
line $\phi=0.46$, solid line $\phi=0.52$, and dashed line
$\phi=0.56$).  (c) The anticorrelation
scale factor $c(\Delta t)$ from Eq.~\protect\ref{back}; see
text for details.
(d) $r_{\rm cage}$ as a function of $\Delta t$.  The symbols
in (b-d) are
the same as part (a).
}
\label{time1}
\end{figure}

\begin{figure}
\centerline{
\epsfxsize=9.0truecm
\epsffile{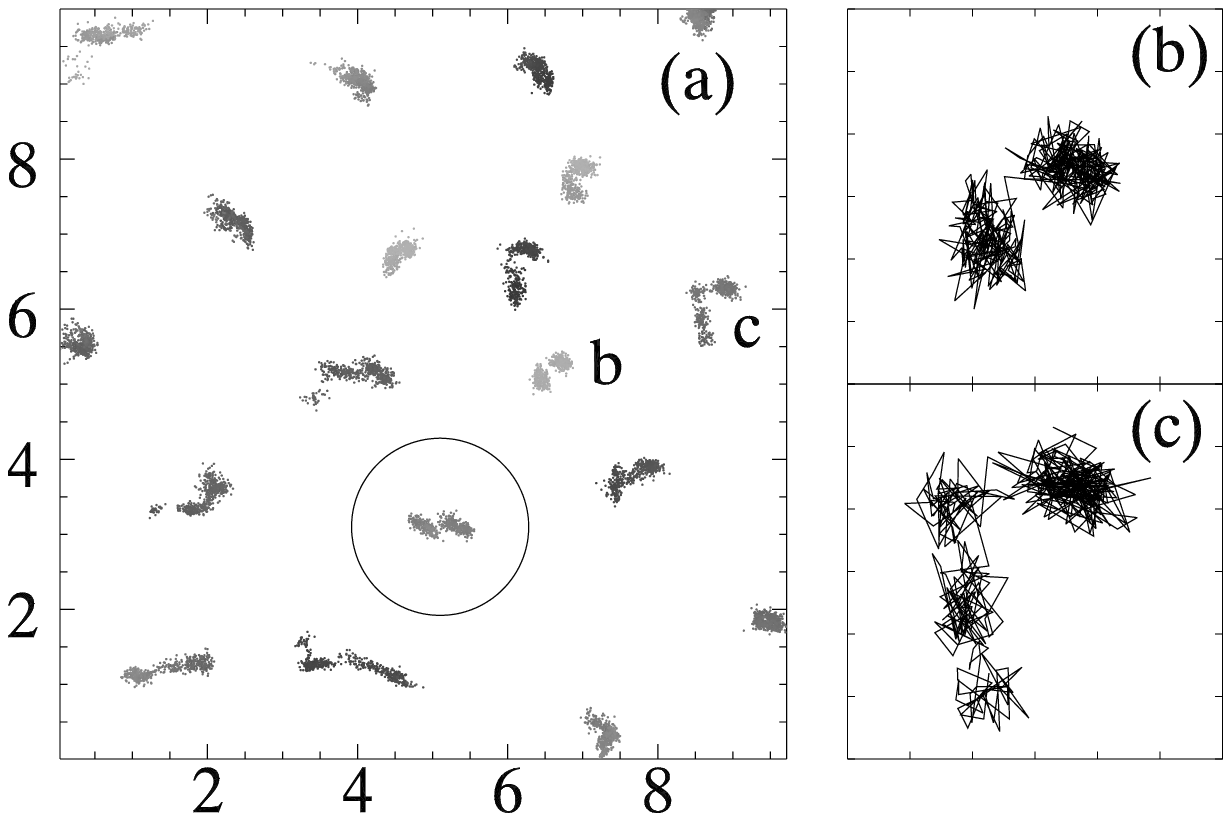}}
\smallskip
\caption{(a) Trajectories of particles from a sample with
$\phi=0.52$, over a 2~hour period.
The axes are labeled in microns, and the
circle illustrates the particle size.
These trajectories are from particles within 
a 2.5~$\mu$m thick region within the
sample; the gray shades indicate vertical distance (darker is
closer to the coverslip).   (b) and (c) are magnifications of
two of the trajectories, with tick marks indicating 0.2~$\mu$m
spacings.  These two particles alternate between being trapped in
a local cage, and a slight jump to a new location when the cage
rearranges.
}
\label{trajs}
\end{figure}

We calculate the mean square displacement $\langle \Delta x^2
\rangle$ from the measured particle positions, and several
typical curves are shown in Fig.~\ref{time1}(a).  The data at short
$\Delta t$ (less than 10~s) is obtained from two-dimensional
measurements within the three dimensional sample, in order
to improve the time resolution.  At the shortest lag times,
$\langle \Delta x^2 \rangle$ increases due to the diffusive
motion of the particles.  At intermediate time scales, the
MSD has a plateau, which becomes more
pronounced as the volume fraction $\phi$ increases toward
$\phi_g \approx 0.58$.  This plateau is due to the cage effect:
particles are trapped in transient cages formed by their
neighbors, and thus cannot diffuse freely through the sample
\cite{reviews,doliwa98}.
At the largest $\Delta t$, the cages rearrange, and particles
are able to diffuse throughout the sample, albeit with a
greatly decreased diffusion coefficient $D_\infty$
\cite{doliwa98,doliwa99,glotzer}.  This can
be seen in the particle trajectories shown in Fig.~\ref{trajs}.
Figure~\ref{trajs}(a) shows two-dimensional projections of
trajectories of several particles within a small region.
The two particles marked {\it b} and {\it c} are magnified
to the right, and show the difference between caged motion,
and the rearrangements.

\begin{figure}
\centerline{
\epsfxsize=6.0truecm
\epsffile{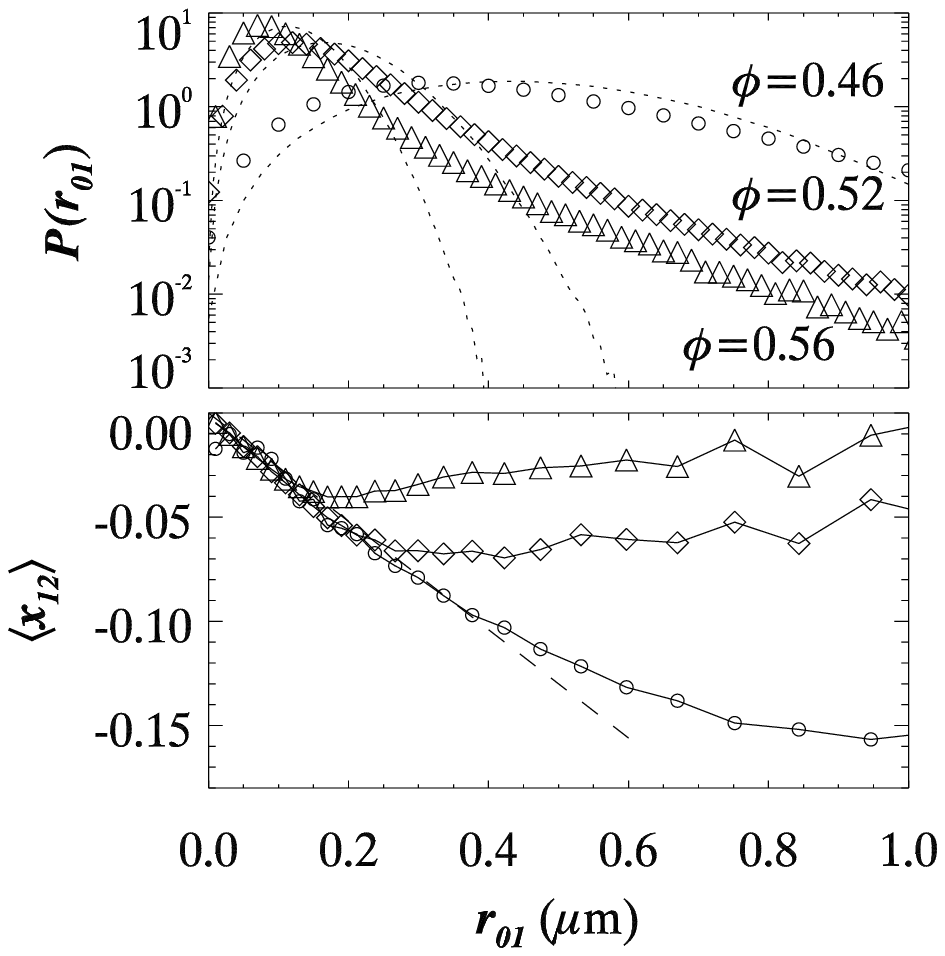}}
\smallskip
\caption{(a) Probability distribution functions for
displacements $r_{01}$ with time scales $\Delta t= 260$~s for
$\phi=0.46$, 700~s for $\phi=0.52$, and 1000~s for $\phi=0.56$.
(b) 
$\langle x_{12} \rangle$ as a function of $r_{01}$
for the same data shown in (a).  
The values of $\Delta t$ for the three data sets have
been chosen to produce similar behavior at small $r_{01}$, which
in these cases is reasonably well described as
$\langle x_{12} \rangle = - (0.26) r_{01}$ (indicated by
the dashed line).  The departure from the small $r_{01}$ behavior
occurs at $r_{\rm cage} \approx 0.75, 0.35, 0.25$~$\mu$m for
$\phi=0.46, 0.52, 0.56$.
}
\label{speed1}
\end{figure}

The cage rearrangements -- that is, the relatively rapid
shifts in particle positions seen in Fig.~\ref{trajs} --
are reflected in broad tails for the distribution of particle
displacements \cite{glotzer,kasper98,marcus99,harrowell96}.
These distributions are shown by the symbols
in Fig.~\ref{speed1}(a).  The time scales for the displacements
are chosen to be comparable to the end of the MSD plateau.
The majority of particles move only
short distances, as they are confined within cages.
However, the distributions show that a nontrivial fraction
of particles do move large distances, more than would be
expected if the distributions of displacements were gaussian
[dotted lines in Fig.~\ref{speed1}(a)].  A traditional way to
quantify the relative size of the tails of the distribution
is to calculate a nongaussian parameter, which compares the
fourth moment of the distribution to the second moment:
\begin{equation}
\alpha_2(\Delta t) = {3 \langle \Delta r^4(\Delta t) \rangle
\over 5 \langle \Delta r^2(\Delta t) \rangle^2} - 1
\end{equation}
which is zero for a gaussian distribution,
and larger when the distribution is broader (for
example, $\alpha_2=1$ for an exponential distribution)
\cite{glotzer,harrowell96,rahman,kob95}.  This parameter is
close to zero at small and large lag times $\Delta t$, and is
a maximum at an intermediate value $\Delta t^*$ which we use to
define the cage rearrangement time scale \cite{doliwa99,glotzer}.  This
time scale is indicated by vertical bars in Fig.~\ref{time1}(a),
and corresponds qualitatively with the end of the MSD plateau.

To quantify the cage effect, we wish to look for temporal
correlations in a particle's motion; we follow the method of
Doliwa and Heuer \cite{doliwa98,doliwa99}.  In particular, if a
particle moves in one direction for a period of time, its
neighboring particles (the ``cage'') will prevent further
motion in that direction, and may push the first particle back
toward the middle of the cage.  In this way, the positions
of the neighboring particles, which have shifted slightly to
allow the interior particle to move, provide a ``memory'' of
the interior particle's motion.  Thus we expect that usually a
particle's motion will be temporally anticorrelated, unless it
is involved in a cage rearrangement (and thus moves and
then stays in its new position).  To look for this, we pick a
time scale $\Delta t$ and then consider displacement vectors
for each particle, $\Delta \vec r_{mn} = \vec r(n\Delta t) -
\vec r(m \Delta t)$.  In particular we wish to determine how
$\Delta \vec r_{12}$ depends on $\Delta \vec r_{01}$, and how
this depends on the time scale $\Delta t$ \cite{doliwa98}.

Anticipating that $\Delta \vec r_{12}$ is directionally
correlated with $\Delta \vec r_{01}$, we consider two components
of $\Delta \vec r_{12}$:  $x_{12}$ is the component of $\Delta
\vec r_{12}$ parallel to $\Delta \vec r_{01}$, the original
displacement, and $y_{12}$ is the component of $\Delta
\vec r_{12}$ along an arbitrarily chosen direction
perpendicular to $\vec r_{01}$ \cite{doliwa98}.  Because of
the arbitrariness in calculating $y_{12}$, 
the average $\langle y_{12} \rangle=0$.
For dilute samples, caging does not occur
and $\langle x_{12} \rangle = 0$;
$\langle x_{12} \rangle$ will be negative if memory effects
are present.
Particles which initially move farther must move their neighbors
farther as well, and so we
expect that $\langle x_{12} \rangle$ will depend on how far
a particle has originally moved, $r_{01} = | \vec r_{01} |$.
To investigate this we compute the average value $\langle
x_{12} \rangle$ as a function of $r_{01}$, and plot this in
Fig.~\ref{speed1}(b) for three different volume fractions
\cite{doliwa98}.  $\Delta t$ has been chosen so that the
curves have similar behavior at small $r_{01}$, and also to
be close to the cage rearrangement time scale $\Delta t^*$.
The average is taken over all particles and all initial times.
$x_{12}$ is negative, indicating anticorrelated motion:
particles which move in one direction during the first time
interval will, on average, move in the opposite direction during
the subsequent time interval.  This is a direct signature of the
cage effect.  Moreover, for particles with small displacements
$r_{01}$, the average subsequent displacement $\langle x_{12}
\rangle$ is linearly proportional to $r_{01}$, as indicated by
the dashed line in Fig.~\ref{speed1}(b).  For larger $r_{01}$,
$\langle x_{12} \rangle$ is no longer proportional to $r_{01}$, and
in fact becomes almost independent of $r_{01}$ \cite{doliwa98}.

The departure from the linear behavior at small $r_{01}$
occurs at smaller distances as the volume fraction $\phi$
increases toward the glass transition.  The existence of two
regimes -- a linear response at small $r_{01}$ and a breakdown
of this linear response at larger $r_{01}$ -- suggests that
the crossover point can be taken as $r_{\rm cage}$, and the
two regimes be identified as caged particles and rearranging
particles respectively.  In other words, particles with $r_{01} <
r_{\rm cage}$ typically remain caged, and the effect of the cage
is to push the particle back toward its original position
\cite{doliwa98,doliwa99}.  The
strength of this effect is given by 
\begin{equation}
\label{back}
\langle x_{12} \rangle = -c r_{01},
\end{equation}
with for example $c=0.26$ for the data shown in
Fig.~\ref{speed1}(b).  Particles with $r_{01} > r_{\rm cage}$
still tend to be pushed back, but not as far as predicted from
linear extrapolation from the small $r_{01}$ behavior:  thus
these particles may end up in new positions, and their
behavior reflects cage rearrangements rather than caged motion.
The changes seen in Fig.~\ref{speed1}(b) as $\phi$ is
increased shows that the cage size $r_{\rm cage}$ decreases as
the glass transition is approached \cite{weeks01,allegrini99}.

By studying the $\Delta t$ dependence of the proportionality
constant $c$ and cage size $r_{\rm cage}$, we can better
understand the MSD.  The value of $c$ depends strongly on the
chosen time scale $\Delta t$, as shown in Fig.~\ref{time1}(c).
In the middle of the MSD plateau, $c$ is large, close to 0.5;
at larger $\Delta t$ it decreases, signaling a diminishing
cage effect.  $c(\Delta t)$ can be related to the logarithmic
slope of the MSD \cite{doliwa98}, to directly connect the cage
effect to the subdiffusive MSD plateau.  Locally the MSD grows as
$\langle \Delta r^2 \rangle \sim \Delta t^{\gamma(\Delta t)}$,
with the anomalous diffusion exponent $\gamma(\Delta t)$ equal
to the logarithmic derivative of $\langle \Delta x^2 \rangle$.
This can be estimated as:
\begin{eqnarray}
\gamma_{\rm est}(\Delta t) &=& {{d \ln \langle \Delta r^2 \rangle} \over {d
\ln \Delta t}}\nonumber\\ 
&\approx&
{\ln [{ | \Delta \vec{r}_{01} + \Delta \vec{r}_{12} |^2
 / \langle r_{01}^2 \rangle}] \over
\ln( { 2\Delta t / \Delta t})}
\nonumber\\
&=& {\ln (2 + 2 \langle x_{12} r_{01} \rangle/\langle r_{01}^2 \rangle) 
\over \ln 2}
\nonumber\\
&\approx &
1 + {\ln(1 - c(\Delta t)) / \ln 2}.
\label{slope}
\end{eqnarray}
We have used $\langle r_{12}^2 \rangle = \langle r_{01}^2
\rangle$ (time invariance) and the final approximation uses
$\langle x_{12} r_{01} \rangle/\langle r_{01}^2 \rangle \approx
\langle x_12 \rangle / \langle r_{01} \rangle \approx
-c$, in analogy with Eq.~\ref{back}; we have verified that
these approximations are reasonable \cite{doliwa99}.  In Fig.~\ref{time1}(b) the
symbols show $\gamma(\Delta t)$ computed directly from the MSD,
and the lines show $\gamma_{\rm est}(\Delta t)$ calculated
from Eq.~\ref{slope}.  The subdiffusive plateau in the MSD is
seen as a broad range of $\Delta t$ for which $\gamma(\Delta t)
< 1$, although it is also clear that $\gamma$ does not have a
constant value anywhere in the plateau, but rather is a smoothly
evolving function of $\Delta t$.  Moreover, the behavior of
$\gamma(\Delta t)$ is well-captured by the calculated value
based on $c(\Delta t)$, as shown by the agreement between
the symbols ($\gamma$) and the lines ($\gamma_{\rm est}$).
In other words, the subdiffusive behavior of the MSD is a
direct consequence of the caged motion of the particles, as
measured by Eq.~\ref{back}.  As $\Delta t$ increases, the cage
effect becomes less important, $c(\Delta t)$ decreases toward
zero (no caging), and the MSD approaches diffusive behavior
($\gamma \rightarrow 1$).

The behavior of the cage size $r_{\rm cage}$ is shown in
Fig.~\ref{time1}(d).  $r_{\rm cage}$ is relatively insensitive
to $\Delta t$, indicating the the size of the cage is more
likely a static property \cite{doliwa98,weeks01}.  The cage
size decreases as the glass transition is approached, although
it has a nonzero value at the glass transition
\cite{weeks01,allegrini99}.
The diffusive behavior of the MSD at large time scales can thus
be thought of as due to the random walks of the individual
particles, each taking steps of size $r_{\rm cage}$ in random
directions \cite{weeks01}.  On the cage rearrangement time
scale $\Delta t^*$, only a few particles move (5 -
10 \%) \cite{weeks00,glotzer,weeks01}, and so in fact the average
time between random walk steps is much larger than $\Delta
t^*$ as seen in Ref.~\cite{weeks01}.


\begin{figure}
\centerline{
\epsfxsize=8.0truecm
\epsffile{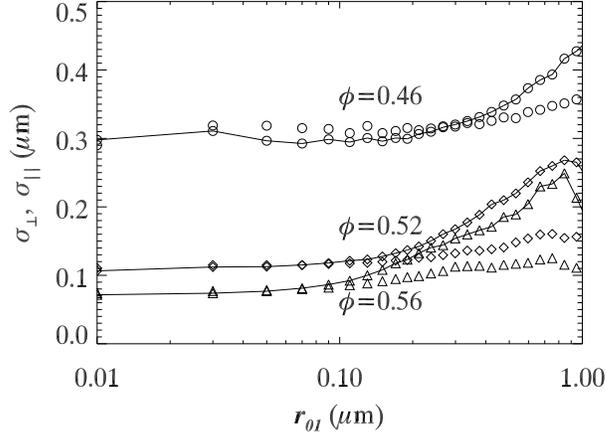}
}
\smallskip
\caption{$\sigma_{\parallel}$ (connected symbols) and $\sigma_{\perp}$
as a function of $r_{01}$, for three different volume fractions
as indicated.  The time scales are as in
Fig.~\protect\ref{speed1}.
}
\label{sigma}
\end{figure}

Further insight into the cage effect can be found by
studying the behavior of the total displacement $\Delta \vec
r_{12}$ rather than focusing only on $x_{12}$, the component
in the direction of $\Delta \vec r_{01}$.
$\Delta \vec r_{12}$ can be decomposed into the deterministic
part
($\langle x_{12} \rangle$ given by Eq.~\ref{back} and $\langle
y_{12} \rangle=0$), and a stochastic
part.  Both the deterministic and stochastic parts may
may depend on $r_{01}$.
To measure the importance of the stochastic part, we
compute $\sigma_{\parallel} = \langle x_{12}^2 \rangle -
\langle x_{12} \rangle^2$ and $\sigma_{\perp} = \langle y_{12}^2
\rangle - \langle y_{12} \rangle^2$, shown
in Fig.~\ref{sigma} by the connected symbols
and unconnected symbols, respectively.  The behaviors of
the parallel and perpendicular components are similar
at small values of $r_{01}$, but differ markedly when
the original displacement has a larger distance $r_{01}$
\cite{doliwa99}.
The transverse component 
$\sigma_{\perp}$ is nearly constant as a function of $r_{01}$,
but $\sigma_{\parallel}$ becomes much larger when $r_{01}$
is larger.  Again, any dependence whatsoever on $r_{01}$
is indicative of memory in the system, and the increase in
$\sigma_{\parallel}$ reflects a memory of {\it mobility}.
Particles which move large distances originally (large
values of $r_{01}$) are more mobile subsequently (large
values of $\sigma_{\parallel}$), and in particular are
more mobile along the direction of the original motion.

\begin{figure}
\centerline{
\epsfxsize=6.0truecm
\epsffile{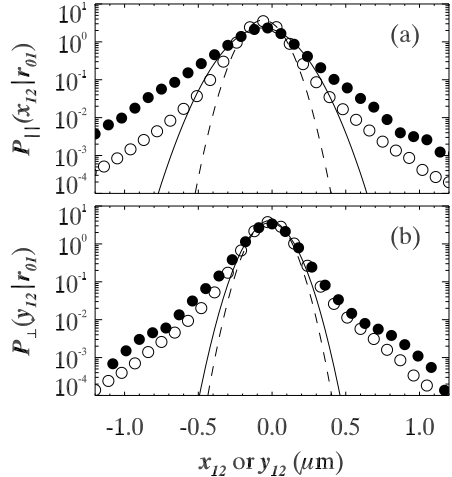}}
\smallskip
\caption{(a) The functions $P_{\parallel}(x_{12} | r_{01}; t)$
and (b) $P_{\perp}(y_{12} | r_{01};t)$, for $\phi=0.52$ (a
liquid), with $t=t^{*} = 600$ s.  The open circles are
all data for $r_{01} < r_{\rm cage}=0.4$~$\mu$m and the closed circles
are for $r_{01} > r_{\rm cage}$.  The gaussian fits are shown as
dashed lines for $r_{01}< r_{\rm cage}$ and solid lines
for $r_{01}> r_{\rm cage}$, and have widths of $\sigma \approx
0.13$~$\mu$m for all except $\sigma_{\parallel}(r_{01} >
r_{\rm cage}=0.22$~$\mu$m.
Similarly, the nongaussian
parameter $\alpha_2 = 1.8$ for all except
$P_{\parallel}(x_{12} | r_{01} > r_{\rm cage})$, which
has $\alpha_2 = 1.0$.
}
\label{pdf}
\end{figure}

Confirmation of this is seen by plotting the
distribution functions $P_{\parallel}(x_{12}|r_{01})$ and
$P_{\perp}(y_{12}|r_{01})$ in Fig.~\ref{pdf}, where the open
circles are for $r_{01} < r_{\rm cage}$ and the closed circles
are for $r_{01} > r_{\rm cage}$.  Gaussian fits to these
distribution functions are shown by the lines.  All of the
functions appear similar, except for $P_{\parallel}(x_{12};
r_{01}>r_{\rm cage}$), which is significantly broader [solid
circles in Fig.~\ref{pdf}(a)].  Thus,
particles which originally have larger displacements are more
likely to continue moving in the same direction (large $x_{01} >
0$) or more likely to move a large distance backwards ($x_{01}
< 0$), but slightly less likely to stay in the same position.
Moreover, as $\sigma_{\parallel} \neq \sigma_{\perp}$, the
particles undergoing cage rearrangements move in a highly
anisotropic fashion.  The distributions are broader than
Gaussians, as can be seen by comparing the symbols to the lines;
this is a reflection of the underlying broad distributions of
the displacements, as shown in Fig.~\ref{speed1}(a).  Note also
that the distributions shown in Fig.~\ref{pdf} are symmetric
about the peak; this is unsurprising for $P_{\perp}(y_{12})$
and perhaps more surprising for $P_{\parallel}(x_{12})$.

\section{Discussion}

We have studied the microscopic motion of thousands of
tracer particles in a concentrated colloidal sample, in
order to understand the dramatic dynamical changes near the
glass transition.  In particular, near the glass transition,
particles are confined to transient cages, resulting in
temporal anticorrelations in particle displacements.  We find
that caging can be described as a deterministic anticorrelated
motion, plus a stochastic part.  The deterministic part is due
to memory provided by the caging particles, which
must adjust their positions to allow a particle to move, and
subsequently push that particle back toward its original position.
By quantifying these effects (as given by Eq.~\ref{back}),
we can connect the properties of the cage directly to the
subdiffusive growth of the mean square displacement (MSD),
shown in Fig.~\ref{time1}(a).  The connection is quite good,
as seen by comparing the lines and symbols in Fig.~\ref{time1}(b).

The long time behavior of the MSD is diffusive, as seen
in Fig.~\ref{time1}(a).  This can be thought of as due
to the random walks taken by the individual particles,
which alternate between being stuck in cages for a random
duration, and a cage rearrangement motion of random length
(see Fig.~\ref{trajs}).  A simple possibility which leads to
diffusive motion at long times is that the cages responsible
for the subdiffusive plateau have finite lifetimes with a
characteristic time scale.  An alternate possibility is that
the cage rearrangement motions could be L\'evy flights.  L\'evy
flights are motions with an infinite mean square step size, in
other words, cage rearrangements would involve movements that
carry particles large distances.  In such a way, diffusive
motion at long times could be due to a competition between
cages with infinite mean lifetime, and motions with infinite
mean square lengths \cite{weeks98}.  (These possibilities would
suggest that the distribution for cage times and/or step sizes
are power laws, for example $P(\Delta x) \sim (\Delta x)^{-\nu}$
for the cage rearrangement displacement $\Delta x$ with $1 <
\nu < 3$.)  L\'evy flights seem possible when looking at the
broad tails shown in Fig.~\ref{speed1}(a).  However, at best
Fig.~\ref{speed1}(a) shows a truncated L\'evy distribution.
We do not see any particles making dramatic displacements
much larger than their own radius; the trajectories shown in
Fig.~\ref{trajs} making small adjustments are typical.  It seems
likelier that the characteristic step size is $r_{\rm cage}$,
a small and finite distance, and thus the diffusive growth of
the MSD as $\Delta t \rightarrow \infty$ is due to a finite
cage lifetime \cite{weeks01}.  In glassy samples, the cage
rearrangements are no longer allowed, and thus the MSD will be
subdiffusive at all times, and perhaps asymptotically reach a
plateau; thus we expect these concepts to be even more useful in
understanding the strange kinetics of nonergodic glassy samples.

We thank B.~Doliwa and H.~G.~E.~Hentschel for helpful
discussions, and thank A.~Schofield for providing our colloidal
samples.  This work was supported by NSF (DMR-9971432) and NASA
(NAG3-2284).


\end{document}